\UseRawInputEncoding
\documentclass[aps,floatfix,preprint,preprintnumbers,nofootinbib,superscriptaddress,natbib]{revtex4}

\usepackage{graphicx,float}
\usepackage{epsfig}		
\usepackage{dcolumn}
\usepackage{bm}
\usepackage{subfigure}

\usepackage[all]{xy}
\usepackage{amsmath,upgreek}
\usepackage{amssymb}

\usepackage{pdfpages}
\usepackage{color}
\usepackage{graphicx,epstopdf}
\usepackage[colorlinks=true,
 linkcolor=red,
 urlcolor=purple,
 citecolor=blue,hyperindex]{hyperref}
\def\0{\mbox{\tiny $0$}}
\def\1{\mbox{\tiny $1$}}
\def\2{\mbox{\tiny $2$}}
\def\3{\mbox{\tiny $3$}}
\def\4{\mbox{\tiny $4$}}
\def\5{\mbox{\tiny $5$}}
\def\6{\mbox{\tiny $6$}}
\def\7{\mbox{\tiny $7$}}
\def\8{\mbox{\tiny $8$}}
\def\9{\mbox{\tiny $9$}}

\def\f14{\mbox{\tiny $\frac{1}{4}$}}

\definecolor{myred}{rgb}{0.9,0,0}
\definecolor{myblue}{rgb}{0,0,0.9}

\begin{document}

\title{Non-commutative phase-space Lotka-Volterra dynamics: the quantum analogue}


\renewcommand{\baselinestretch}{1.2}
\author{A. E. Bernardini}
\email{alexeb@ufscar.br}
\altaffiliation[On leave of absence from]{~Departamento de F\'{\i}sica, Universidade Federal de S\~ao Carlos, PO Box 676, 13565-905, S\~ao Carlos, SP, Brasil.}
\author{O. Bertolami}
\email{orfeu.bertolami@fc.up.pt}
\altaffiliation[Also at~]{Centro de F\'isica das Universidades do Minho e do Porto, Rua do Campo Alegre s/n, 4169-007, Porto, Portugal.} 
\affiliation{Departamento de F\'isica e Astronomia, Faculdade de Ci\^{e}ncias da
Universidade do Porto, Rua do Campo Alegre 687, 4169-007, Porto, Portugal.}
\date{\today}

\begin{abstract}
The Lotka-Volterra (LV) dynamics is investigated in the framework of the Weyl-Wigner (WW) quantum mechanics (QM) extended to one-dimensional Hamiltonian systems, $\mathcal{H}(x,\,k)$, constrained by the $\partial^2 \mathcal{H} / \partial x \, \partial k = 0$ condition. Supported by the Heisenberg-Weyl non-commutative algebra, where $[x,\,k] = i$, the canonical variables $x$ and $k$ are interpreted in terms of the LV variables, $y = e^{-x}$ and $z = e^{-k}$, eventually associated with the number of individuals in a closed competitive dynamics: the so-called prey-predator system. The WW framework provides the ground for identifying how classical and quantum evolution coexist at different scales, and for quantifying {\it quantum analogue} effects. 
Through the results from the associated Wigner currents, (non-)Liouvillian and stationary properties are described for thermodynamic and gaussian quantum ensembles in order to account for the corrections due to quantum features over the classical phase-space pattern yielded by the Hamiltonian description of the LV dynamics.
In particular, for gaussian statistical ensembles, the Wigner flow framework provides the exact profile for the quantum modifications over the classical LV phase-space trajectories so that gaussian quantum ensembles can be interpreted as an adequate Hilbert space state configuration for comparing quantum and classical regimes.
The generality of the framework developed here extends the boundaries of the understanding of quantum-like effects on competitive microscopical bio-systems.
\end{abstract}


\date{\today}
\maketitle

\section{Introduction}

The Lotka-Volterra (LV) \cite{LV} dynamical systems, often described by a nonlinear Hamiltonian, $\mathcal{H}(x,\,k)$, constrained by the condition $\partial^2 \mathcal{H} / \partial x \, \partial k = 0$, have been considered in a wide range of effective models, which include the description of stability criteria for complex communities in microbiology \cite{Nature01, Nature02}, phase transitions in finite microbiological systems \cite {PRE-LV3}, and also extinction divergence (or even immigration) effects \cite{SciRep02,PRE-LV2}. 
Furthermore, through the Fokker-Planck equation, effects of fluctuations around (classical) stability conditions described by a mean-field approach have already been quantified \cite{Agui01} and, interestingly, considered for interpreting prey-predator system instabilities in terms of the phase-space Weyl-Wigner (WW) quantum mechanics (QM) \cite{PRE-LV}. In this context, quantum fluctuations could be identified, quantified, and eventually re-interpreted \cite{NossoPaper,Novo2022} to map some collective behavior exhibited by biological systems.

\begin{figure}[h]
\includegraphics[scale=0.6]{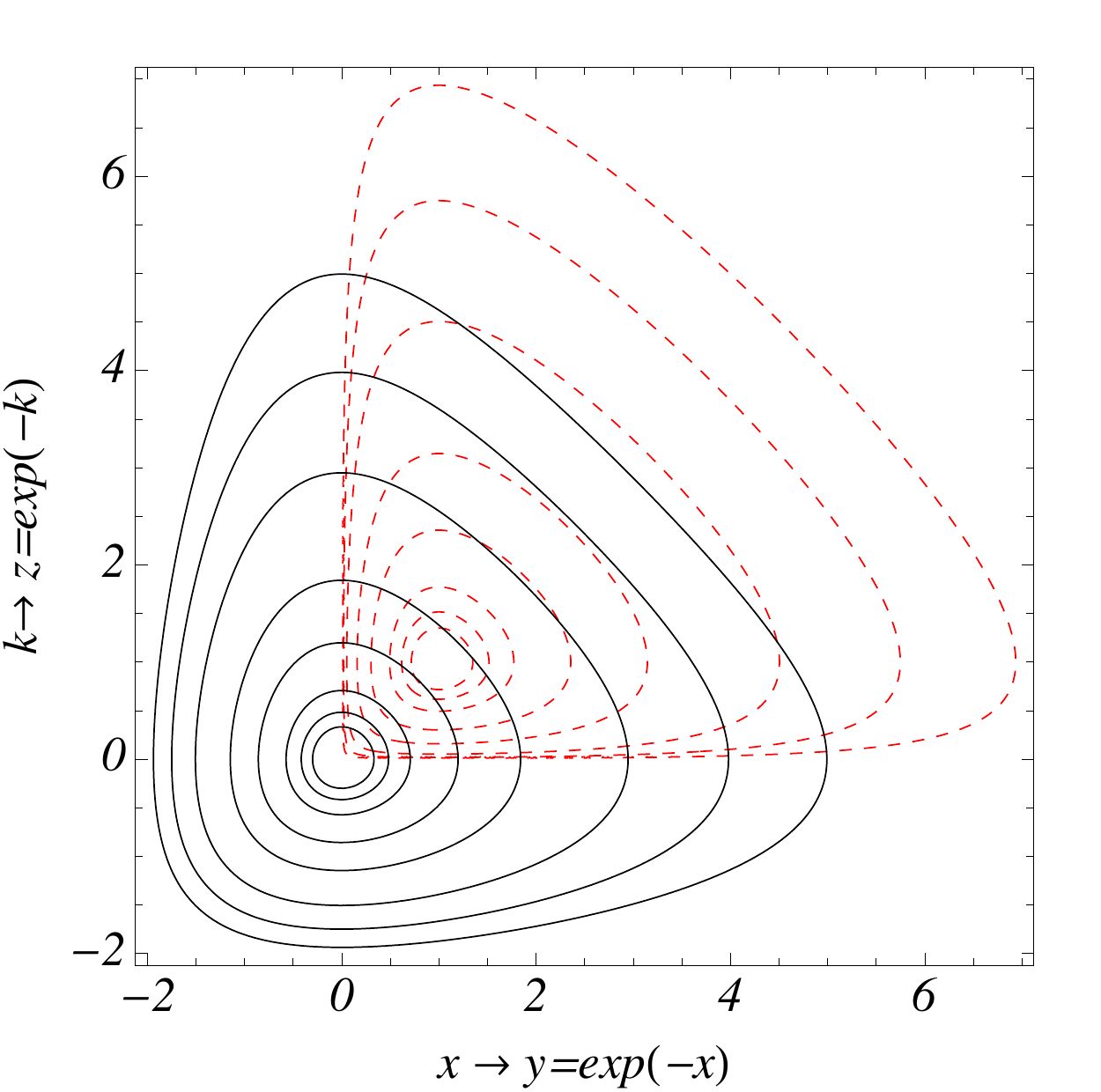}
\renewcommand{\baselinestretch}{.65}
\caption{\footnotesize
(Color online) Classical portrait of Lotka-Volterra associated Hamiltonians. Phase-space $x-k$ trajectories (black solid lines) for $\epsilon = 6,\, 5,\, 4,\, 3,\, 2.5,\, 2.2,\, 2.1$ and $2.05$, and corresponding species distributions (red dashed lines), $z$ and $y$, identified by $y = e^{-x}$ and $z = e^{-k}$.}
\label{LVLV}
\end{figure}

Besides the straightforward description of competitive ecological equilibrium of populations \cite{RPSA-LV,Anna}, LV systems are currently considered in the investigation of stochastic systems \cite{Allen,Grasman}, and even in broader contexts, in which their Hamiltonian structure is reconfigured, for instance, for describing the cosmological dynamics \cite{SciRep01}.
In fact, LV systems exhibit subtle mathematical features which can be encompassed by a dimensionless Hamiltonian described by 
\begin{equation}\label{Ham}
\mathcal{H}(x,\,k) = a \,x + k + a\, e^{-x} + e^{-k},
\end{equation}
resulting into classical equations of motion,
\begin{eqnarray}\label{Ham2}
d{x}/d\tau &=& \{x,\mathcal{H}\}_{PB} = 1-e^{-k},\nonumber\\
d{k}/d\tau &=& \{k,\mathcal{H}\}_{PB} = a\,e^{-x} - a,
\end{eqnarray}
which describe $x-k$ oscillations as a natural mode of population coexistence in ecological chains, with the number of individuals, $y$ and $z$, identified by $y = e^{-x}$ and $z = e^{-k}$, such that
\begin{eqnarray}\label{Hamss}
d{y}/d\tau &=& z\, y - y,\nonumber\\
d{z}/d\tau &=& a(z -z \,y).
\end{eqnarray}
In particular, for the isotropic coordinate condition, $a = 1$, phase-space trajectories as depicted in Fig.~\ref{LVLV}, are parameterized by an auxiliary variable, $\mathcal{T}$, and a dimensionless energy parameter, $\epsilon$, as
\begin{eqnarray}\label{Ham2B}
x(\mathcal{T}) &=& \ln(2) - \ln\left[\mathcal{T} \pm \sqrt{\mathcal{T}^2-4 e^{\mathcal{T} - \epsilon}}\right]\quad\Rightarrow \quad y(\mathcal{T})=\frac{\mathcal{T} \pm \sqrt{\mathcal{T}^2-4 e^{\mathcal{T} - \epsilon}}}{2},\nonumber\\
k(\mathcal{T}) &=& \ln(2) - \ln\left[\mathcal{T} \mp \sqrt{\mathcal{T}^2-4 e^{\mathcal{T} - \epsilon}}\right]\quad\Rightarrow \quad z(\mathcal{T})=\frac{\mathcal{T} \mp \sqrt{\mathcal{T}^2-4 e^{\mathcal{T} - \epsilon}}}{2},
\end{eqnarray}
subjected to the constraint,
\begin{eqnarray}\label{Ham2C}
\dot{\mathcal{T}}^2 - {\mathcal{T}^2 + 4 e^{\mathcal{T} - \epsilon}}=0,
\end{eqnarray}
where ``{\it dots}'' correspond to time ($\tau$) derivatives, $d/d\tau$, and $\mathcal{H}(x,\,k) = \epsilon$ is identified as a classical phase-space trajectory.

The mathematical modeling of the above dynamics indeed allows for better understanding of ecosystem-like interactions, such as, for instance, competition, symbiosis and species extinction \cite{PRE-LV,PRE-LV2,PRE-LV3,PRE-LV4}, as well as of macroscopic effects like stable coexistence and oscillations.
Such effects can have more complex origins as, for instance, those observed in molecular systems {\it in vitro} \cite{Nature01,PP01}. Prey-predator-like oscillations, competition-induced chaos, and symbiotic synchronization are unprecedented examples of such microscopic behavior. These have been experimentally detected in the study of biochemical and genomic systems \cite{PP00,PP02,PP03,PP04} and can then be regarded as a motivation for considering the quantization of LV systems.

Therefore, it is quite relevant to understand how classical macroscopic and quantum microscopic evolution coexist at different scales, and how quantum effects arise at measurable scales.
Of course, this includes some fundamental questions, such as how to systematically introduce quantum effects into the classical LV dynamics, and if it is possible to evaluate and interpret its consequences as basic features of classical living organisms or biochemical systems.

Likewise, the quantization paradigm may involve some assumptions for the understanding of classical collective behavior \cite{Bio17}.
More specifically, an explanation for time evolution and space state configurations of species distributions, including the measurement procedure over which a {\it quantum analog} non-commutative property of quantum observables $x$ and $k$, with $[x,\,k]\neq 0$, could be evinced and considered.
This includes the definition of minimal phase-space elementary volume, $\delta x \, \delta k > 0$, and corresponds to the situation in which an individual belonging to a species distribution is interpreted as a quantum state, and that the predictability associated with species existence would be associated with the probability amplitude for the state detection.

For addressing these questions, our work reports about the WW formalism \cite{Wigner,Ballentine,Case} 
extended to generalized non-linear Hamiltonians described by $\mathcal{H}(x,\,k) = \mathcal{K}(k) + \mathcal{V}(x)$, through which the bridge between classical Liouvillian and quantum non-Liouvillian frameworks is straightforwardly accessed \cite{Novo2022,NossoPaper,Meu2018}. 

Through the WW representation of QM, it is possible to quantify phase-space correlations and information flow aspects as well as quantum-like deviations from classical patterns. 
As a phase-space extended equivalent description of the Schr\"odinger QM, the WW framework encompasses the QM paradigms in terms of a {\it quasi}-probability distribution function of position and momentum coordinates \cite{Wigner,Hillery,Ballentine,Case}.
Phase-space features of the Wigner flow for generic one-dimensional Hamiltonian systems like the one driven by Eq.~(\ref{Ham}) are naturally engendered by associated Wigner functions and Wigner currents \cite{Novo2022}.
It has already been shown that Liouvillian and stationary properties can be analytically computed for associated thermodynamic (TD) and gaussian quantum ensembles in order to account for the quantum distortions over classical phase-space trajectories \cite{Hillery,Zurek01,Zurek02,Steuernagel3}.
Consequently, once such a statistical description is provided, the Wigner flow quantum features, described in terms of the corresponding Wigner currents, can be related to quantum fluctuations over the classical regime. Once specialized to LV systems, classicality and quantumness can be quantified through probability and information fluxes from a classical Hamiltonian background. 
These tools allows for a systematic analysis of the non-linear LV dynamics encompassed in the framework of the WW QM.

The outline of our manuscript is thus as follows.
Section II is concerned with the WW framework, where the quantum distortion quantifiers related to the Hamiltonian dynamics described by $\mathcal{H}(x,\,k) = \mathcal{K}(k) + \mathcal{V}(x)$ are presented \cite{Novo2022}.
In Section III, the results are specialized to TD ensembles supported by the LV Hamiltonian, Eq.~(\ref{Ham}), which admits a precise interpretation of stationarity conditions in the context of the classical-to-quantum correspondence, providing a connection with statistical mechanics.
A more accurate analysis is developed for gaussian ensembles in Section IV. Again, once specialized to the LV Hamiltonian from Eq.~(\ref{Ham}), the Wigner flow framework provides the overall quantum distortion over the phase-space classical pattern, where the quantum effects are analytically computed through a convergent infinite series expansion in terms of quadratic powers of the Planck constant, $\hbar$. 
The result accounts for the exact quantum corrections over LV classical profile, establishing a natural correspondence between classical and quantum scenarios in terms of quantum correlation quantifiers obtained through the Wigner currents.
All the results are obtained in terms of analytical expressions for the Wigner currents for both TD and gaussian ensembles.
In particular, for gaussian ensembles, it is possible to identify how quantum corrections would affect extinction and stability scenarios related to prey-predator analogue systems \cite{PRE-LV,PRE-LV2,PRE-LV3,PRE-LV4}.
Finally, a subtle aspect involving the classical-quantum correspondence for a highly non-linear Hamiltonian system similar to the LV one, which allows for identifying a kind of phase-space {\it camouflage} of quantum distortions coupled to non-linear effects, is discussed at the end of Section IV.
Our conclusions are presented in Section V.

\section{Quantum Weyl-Wigner phase-space formulation}

Based on the one-dimensional Heisenberg-Weyl algebra, depicted by the position-momentum non-commutative relation, $[\hat{q},\,\hat{p}] = i\,\hbar$, the Wigner {\it quasi}-distribution function, $W(q,\, p)$, extends the QM framework to the phase-space.
From the Weyl transform of an operator $\hat{O}(\hat{q}, \hat{p})$,
\begin{equation}
O^W(q,\, p)\label{111}
= 2\hspace{-.1cm} \int^{+\infty}_{-\infty} \hspace{-.35cm}ds\,\exp{\left[2\,i \,p\, s/\hbar\right]}\,\langle q - s | \hat{O} | q + s \rangle=2\hspace{-.1cm} \int^{+\infty}_{-\infty} \hspace{-.35cm} dr \,\exp{\left[-2\, i \,q\, r/\hbar\right]}\,\langle p - r | \hat{O} | p + r\rangle,
\end{equation} 
if the position and momentum operators, $\hat{q}$ and $\hat{p}$, are converted into $c$-numbers, $q$ and $p$, the Wigner function is straightforwardly identified with the density matrix operator, $\hat{\rho} = |\psi \rangle \langle \psi |$, through the overlap integral given by
\begin{equation}
 (2\pi \hbar)^{-1} \hat{\rho} \to W(q,\, p) = (\pi\hbar)^{-1} 
\int^{+\infty}_{-\infty} \hspace{-.35cm}ds\,\exp{\left[2\, i \, p \,s/\hbar\right]}\,
\psi(q - s)\,\psi^{\ast}(q + s),\label{222}
\end{equation}
which can also be identified with the Fourier transform of the off-diagonal elements of $\hat{\rho}$.
Akin to the formalism of statistical mechanics, the WW phase-space formulation accounts for quantum corrections to TD equilibrium states \cite{Wigner}, and more generically, for all the QM features which are recovered from its marginal distributions obtained through integrations over position and momentum coordinates\footnote{Given by, respectively,
\begin{equation}
\vert \psi_q(q)\vert^2 = \int^{+\infty}_{-\infty} \hspace{-.35cm}dp\,W(q,\, p)
\qquad
\leftrightarrow
\qquad
\vert \psi_p(p)\vert^2 = \int^{+\infty}_{-\infty} \hspace{-.35cm}dq\,W(q,\, p),
\end{equation}
with
\begin{equation}
\psi_q(q) =
(2\pi\hbar)^{-1/2}\int^{+\infty}_{-\infty} \hspace{-.35cm} dp\,\exp{\left[-i \, p \,q/\hbar\right]}\,
\psi_p(p).
\end{equation}}.
Regarding the most fundamental property of the Weyl transform, the trace of the product of two operators, $\hat{O}_1$ and $\hat{O}_2$, evaluated through an overlap integral over the infinite volume described by the phase-space coordinates \cite{Wigner,Case}, $q$ and $p$, 
\begin{equation}
Tr_{\{q,p\}}\left[\hat{O}_1\hat{O}_2\right] = h^{-1} 
\int \hspace{-.15cm}\int \hspace{-.15cm} dq\,dp \,O^W_1(q,\, p)\,O^W_2(q,\, p),
\end{equation}
allows for computing the averaged values of quantum observables, $\hat{O}$, in terms of the Weyl products in the form of
\begin{equation}
 \langle O \rangle = 
\int^{+\infty}_{-\infty} \hspace{-.35cm}dp\int^{+\infty}_{-\infty} \hspace{-.35cm} {dq}\,\,W(q,\, p)\,{O^W}(q,\, p), \label{eqfive}
\end{equation}
which reflects the trace probability properties involving $\hat{\rho}$ and $\hat{O}$, $Tr_{\{q,p\}}\left[\hat{\rho}\hat{O}\right]$. In fact, it assumes a consistent QM interpretation constrained by the normalization condition over $\hat{\rho}$, as $Tr_{\{q,p\}}[\hat{\rho}]=1$\footnote{Such a definition also admits an statistical QM extension, from pure states to statistical mixtures, with the quantum purity expressed in terms of the trace operation, 
\begin{equation}
Tr_{\{q,p\}}[\hat{\rho}^2] = 2\pi\hbar\int^{+\infty}_{-\infty}\hspace{-.35cm}dp\int^{+\infty}_{-\infty} \hspace{-.35cm} {dq}\,\,\,W(q,\, p)^2,
\label{eqpureza}
\end{equation}
built from the replacement of ${O^W}(q,\, p)$ by $W(q,\, p)$ into Eq.~\eqref{eqfive}, and which exhibits the pure state constraint, $Tr_{\{q,p\}}[\hat{\rho}^2] = Tr_{\{q,p\}}[\hat{\rho}] = 1$.}.

The time evolution of the Wigner function, $W(q,\,p) \to W(q,\,p;\,t)$, is constrained by the continuity equation \cite{Case,Ballentine,Steuernagel3,NossoPaper,Meu2018},
\begin{equation}
{\partial_t W} + {\partial_q J_q}+{\partial_p J_p} =0,\qquad \mbox{with}\quad \partial_a \equiv \partial/\partial a,
\label{z51}
\end{equation}
which is associated with the Hamiltonian dynamics driven by a Hamiltonian operator, 
\begin{equation}
{H}(\hat{Q},\,\hat{P}) = K(\hat{P}) + V(\hat{Q}),
\end{equation}
for which the Weyl transform corresponds to ${H}^W(q,\,p) = K(p) + V(q)$.
The flow field connected to the Wigner function dynamics is described in terms of a vector flux, $\mathbf{J} = J_q\,\hat{q} + J_p\,\hat{p}$, with the Wigner current components given by \cite{Steuernagel3,NossoPaper,Meu2018,Novo2022}
\begin{equation}
J_q(q,\,p;\,t) = +\sum_{\eta=0}^{\infty} \left(\frac{i\,\hbar}{2}\right)^{2\eta}\frac{1}{(2\eta+1)!} \, \left[\partial_p^{2\eta+1} K(p)\right]\,\partial_q^{2\eta}W(q,\,p;\,t),
\label{z500BB}
\end{equation}
and
\begin{equation}
J_p(q,\,p;\,t) = -\sum_{\eta=0}^{\infty} \left(\frac{i\,\hbar}{2}\right)^{2\eta}\frac{1}{(2\eta+1)!} \, \left[\partial_q^{2\eta+1} V(q)\right]\,\partial_p^{2\eta}W(q,\,p;\,t),\label{z500CC}
\end{equation}
from which one notices that the quantum effects on the evolution of $W(q,\,p;\,t)$ are due to the series expansion contributions for $\eta \geq 1$. In fact, once these effects are suppressed, the classical Liouvillian regime is recovered and given by
\begin{equation}
J^{\mathcal{C}}_q(q,\,p;\,t)= +({\partial_p H})\,W(q,\,p;\,t), \label{z500BB2}
\end{equation}
and
\begin{equation}
J^{\mathcal{C}}_p(q,\,p;\,t) = -({\partial_q H})\,W(q,\,p;\,t),
\label{z500CC2}
\end{equation}
with the upper index ``$W$'' for $H$ being suppressed from this point onwards.

Now, in order to properly connect the above framework to the LV formulation given in terms of dimensionless phase-space coordinates, $x$ and $k$, one should recast the quantities in terms of dimensionless variables: $x = \left(m\,\omega\,\hbar^{-1}\right)^{1/2} q$ and $k = \left(m\,\omega\,\hbar\right)^{-1/2}p$, where $m$ is a mass scale parameter and $\omega$ is an arbitrary angular frequency. The Hamiltonian Eq.~\eqref{Ham} takes then the form of $\mathcal{H}(x,\,k) = \mathcal{K}(k) + \mathcal{V}(x)$, where $\mathcal{H} = (\hbar \omega)^{-1} H$, with $\mathcal{V}(x) = (\hbar \omega)^{-1} V\left(\left(m\,\omega\,\hbar^{-1}\right)^{-1/2}x\right)$ and $\mathcal{K}(k) = (\hbar \omega)^{-1} K\left(\left(m\,\omega\,\hbar\right)^{1/2}k\right)$. In addition, the Wigner function can be recast in the form of $\mathcal{W}(x, \, k;\,\tau) \equiv \hbar\, W(q,\,p;\,t)$, with
\begin{eqnarray}\label{xDimW}
\mathcal{W}(x, \, k;\,\tau) &=& \pi^{-1} \int^{+\infty}_{-\infty} \hspace{-.35cm}dy\,\exp{\left(2\, i \, k \,y\right)}\,\phi(x - y;\,\tau)\,\phi^{\ast}(x + y;\,\tau),
\end{eqnarray}
and with $\hbar$ absorbed by $dp\,dq\to \hbar\, dx\,dk$ integrations, $y = \left(m\,\omega\,\hbar^{-1}\right)^{1/2} s$, and $\tau = \omega t$, such that Wigner currents are cast into the form of \cite{Novo2022}
\begin{eqnarray}\label{alexDimW}
\label{imWA}\mathcal{J}_x(x, \, k;\,\tau) &=& +\sum_{\eta=0}^{\infty} \left(\frac{i}{2}\right)^{2\eta}\frac{1}{(2\eta+1)!} \, \left[\partial_k^{2\eta+1}\mathcal{K}(k)\right]\,\partial_x^{2\eta}\mathcal{W}(x, \, k;\,\tau),\\
\label{imWB}\mathcal{J}_k(x, \, k;\,\tau) &=& -\sum_{\eta=0}^{\infty} \left(\frac{i}{2}\right)^{2\eta}\frac{1}{(2\eta+1)!} \, \left[\partial_x^{2\eta+1}\mathcal{V}(x)\right]\,\partial_k^{2\eta}\mathcal{W}(x, \, k;\,\tau),
\end{eqnarray}
which lead to the dimensionless continuity equation written as 
\begin{equation}\label{z51dim}
{\partial_{\tau} \mathcal{W}} + {\partial_x \mathcal{J}_x}+{\partial_k \mathcal{J}_k} = {\partial_{\tau} \mathcal{W}} + \mbox{\boldmath $\nabla$}_{\xi}\cdot\mbox{\boldmath $\mathcal{J}$} =0,
\end{equation}
with the explicit form of the stationarity quantifier given by \cite{Novo2022}
\begin{equation} \label{helps}
\partial_{\tau} \mathcal{W} = \sum_{\eta=0}^{\infty}\frac{(-1)^{\eta}}{2^{2\eta}(2\eta+1)!} \, \left\{
\left[\partial_x^{2\eta+1}\mathcal{V}(x)\right]\,\partial_k^{2\eta+1}\mathcal{W}
-
\left[\partial_k^{2\eta+1}\mathcal{K}(k)\right]\,\partial_x^{2\eta+1}\mathcal{W}
\right\},\end{equation}
where the contributions from quantum corrections ($\mathcal{O}(\hbar^{2\eta})$) have been given in terms of the re-defined phase-space coordinates, $\mbox{\boldmath $\xi$} = (x,\,k)$.

More importantly, the interpretation of the quantum effects as fluctuations over the classical regime provides a more accurate meaning to the Liouvillianity \cite{NossoPaper,Novo2022}.
Eq.~\eqref{z51dim} with Wigner currents replaced by the dimensionless forms from Eqs.~\eqref{z500BB2}-\eqref{z500CC2}
results in the classical Liouville equation, with the dimensionless phase-space velocity along a classical trajectory $\mathcal{C}$ identified by $\mathbf{v}_{\xi(\mathcal{C})} = \dot{\mbox{\boldmath $\xi$}} = (\dot{x},\,\dot{k})\equiv ({\partial_k \mathcal{H}},\,-{\partial_x \mathcal{H}})$, which exhibits the divergenceless behavior, $\mbox{\boldmath $\nabla$}_{\xi}\cdot \mathbf{v}_{\xi(\mathcal{C})}= \partial_x \dot{x} + \partial_k\dot{k} = 0$, typical of the Liouvillian regime \cite{NossoPaper,Novo2022}.
In order to extend such a definition to the quantum regime, a quantum current associated velocity is implicitly parameterized by, $\mbox{\boldmath $\mathcal{J}$} = \mathbf{w}\,\mathcal{W}$, with $\mathbf{w}$ corresponding to the {\it quantum analog} of $\mathbf{v}_{\xi(\mathcal{C})}$.
Hence, $\mbox{\boldmath $\nabla$}_{\xi}\cdot\mbox{\boldmath $\mathcal{J}$} = \mathcal{W}\,\mbox{\boldmath $\nabla$}_{\xi}\cdot\mathbf{w}+ \mathbf{w}\cdot \mbox{\boldmath $\nabla$}_{\xi}\mathcal{W}$ can be recast as \cite{Steuernagel3}
\begin{equation}
\mbox{\boldmath $\nabla$}_{\xi} \cdot \mathbf{w} = \frac{\mathcal{W}\, \mbox{\boldmath $\nabla$}_{\xi}\cdot \mbox{\boldmath $\mathcal{J}$} - \mbox{\boldmath $\mathcal{J}$}\cdot\mbox{\boldmath $\nabla$}_{\xi}\mathcal{W}}{\mathcal{W}^2},
\label{zeqnz59}
\end{equation}
from which any local discrepancy from the classical regime is identified by $\mbox{\boldmath $\nabla$}_{\xi} \cdot \mathbf{w}\neq 0$. The Liouvillianity quantifier (cf. Eq.~\eqref{zeqnz59}) is then given by \cite{Novo2022}
\begin{equation}
\mbox{\boldmath $\nabla$}_{\xi} \cdot \mathbf{w} = \sum_{\eta=0}^{\infty}\frac{(-1)^{\eta}}{2^{2\eta}(2\eta+1)!}
\left\{
\left[\partial_k^{2\eta+1}\mathcal{K}(k)\right]\,
\partial_x\left[\frac{1}{\mathcal{W}}\partial_x^{2\eta}\mathcal{W}\right]
-
\left[\partial_x^{2\eta+1}\mathcal{V}(x)\right]\,
\partial_k\left[\frac{1}{\mathcal{W}}\partial_k^{2\eta}\mathcal{W}\right]
\right\}. ~~~\end{equation}

The above quantifier can be considered as a measure of classicality. Nevertheless, it is worth to mention that further analysis of classical to quantum transition aspects have already been extensively considered \cite{Berry79}.
In particular, expecting to distinguish integrable from chaotic motion in (phase-space) classical mechanics, and to evaluate the discrepancies between classical and quantum descriptions for non-linear systems, the Weyl representation connected to new generating functions described by chords and centers in phase-space \cite{DeAlmeida98} were evaluated, even for more sophisticated topologies where the phase-space symplectic area (or action) is bounded a torus and the torus chord is centered at position coordinates \cite{Hannay82}.
In this case, both classical and quantum theories are related by the group of translations and reflections through a point in phase-space.
However, although the Hamiltonian version of the LV system is highly nonlinear, the quantifiers for stationarity and classicality regimes discussed above, $\mbox{\boldmath $\nabla$}_{\xi}\cdot\mbox{\boldmath $\mathcal{J}$}$ and $\mbox{\boldmath $\nabla$}_{\xi} \cdot \mathbf{w}$, suffice for obtaining the information content for distinguishing quantum from classical regimes as well as for identifying their corresponding stability conditions.

Turning back to the Hamiltonian Eq.~\eqref{Ham}, the evolution of ensembles of species $y = e^{-x}$ and $z = e^{-k}$ results in stable (stationary) classical (Liouvillian) patterns corresponding to probability-like distribution contours shown in Fig.~\ref{LVLV}, which can be reproduced by classical Wigner currents for which
$\mbox{\boldmath $\nabla$}_{\xi} \cdot \mbox{\boldmath $\mathcal{J}$} = 0$ and $\mbox{\boldmath $\nabla$}_{\xi} \cdot \mathbf{w} = 0$.
In this case, $x$ and $k$ commute, i.e. $[x,\,k] = 0$, meaning that simultaneous measurements of $x$ and $k$ has no additional constraint(s) besides the Hamiltonian one, $\mathcal{H}(x,\, k)= \epsilon$. Moreover, given that the fluctuations of the number of species parameterized by $\delta x$ and $\delta k$ follow a deterministic evolution parameterized by Eqs.~\eqref{Ham2B} and \eqref{Ham2C}, one has $\delta x\,\delta k = 0$. If on one hand, this can be understood as an {\it ab initio} extinction hypothesis, it is possible, on the other hand, to assume a {\it quantum analog} hypothesis from which either $[x,\,k] = i$ or, in a softer version, $[x,\,k] \neq 0$, meaning that measurements of $x$ and $k$ affect one each other, and therefore cannot be performed simultaneously. The {\it quantum analogs} of the uncertainty principle are expressed either by $\delta x\,\delta k \gtrsim 1$ or by $\delta x\,\delta k \gtrsim 0$, corresponding to either $[x,\,k] = i$ or $[x,\,k] \neq 0$, respectively.
For this quantized version of $x$ and $k$ coordinates, the fluctuations of the number of species, $\delta x$ and $\delta k$, follow a non-deterministic evolution parameterized by Wigner currents, which drive the statistical and probability behaviors over the quantum ensembles. Therefore, it can be interpreted as a suppression of an extinction outcome given that it precludes $\delta x\,\delta k = 0$. Of course, even exhibiting a non-Liouvillian pattern, with $\mbox{\boldmath $\nabla$}_{\xi} \cdot \mathbf{w} \neq 0$, the {\it quantum analog} regime also admits stable evolution patterns, where $\mbox{\boldmath $\nabla$}_{\xi} \cdot \mbox{\boldmath $\mathcal{J}$} = 0$. This will be discussed in the following sections, where statistical ensembles are considered.

\section{Thermodynamic ensembles}

The Maxwell-Boltzmann distribution for classical statistical ensembles is given by
\begin{equation}\label{TDclass}
\mathcal{W}_0(x,\,k;\,\beta) = \mathcal{Z}^{-1}_0(\beta)\,\exp[-\beta\,\mathcal{H}(x,\,k)],
\end{equation}
with the corresponding partition function,
\begin{equation}
\mathcal{Z}_0(\beta) = 
\int^{+\infty}_{-\infty} \hspace{-.35cm}dx\,\int^{+\infty}_{-\infty} \hspace{-.35cm}dk\,\exp[-\beta\,\mathcal{H}(x,\,k)],
\end{equation}
where the dependence on the temperature, $T$, is introduced through the dimensionless parameter $\beta = \hbar \omega/ k_{B} T$, where $k_{B}$ is the Boltzmann constant.
The quantum analog of $\mathcal{W}_0(x,\,k;\,\beta)$ is given by $\sum_{\ell}\exp(- E_{\ell}/k_BT) \mathcal{W}_{\ell}(x,\, k)$ where $\ell$ is the quantum number for {\it eigen}energies, $E_{\ell}$, and Weyl-Wigner transformed stationary {\it eigen}functions, $\mathcal{W}_{\ell}(x,\, k)$.
Even though the quantum system spectral decomposition, $\{E_{\ell}\}$, cannot be identified, quantum corrections to the classical TD regime can be obtained through a perturbative procedure.
In fact, stationary solutions for the Wigner continuity equation (cf. Eq.~\eqref{z51dim}), $\mathcal{W}_{ST}(x,\, k;\,\beta)$, can be iteratively obtained through the dimensionless series expansion given by \cite{Wigner,Coffey07,Pol2010,Hillery,Zurek03}\footnote{It is given by \begin{equation} 
{W}^{(2N)}_{ST}(q,\, p;\,\beta) = \sum_{\eta=0}^N\,\hbar^{2\eta}\,{W}_{2\eta}(q,\, p;\,\beta),\nonumber
\end{equation}
if one expresses it in position and momentum physical coordinates, $q$ and $p$ \cite{Wigner,Coffey07}.}
\begin{equation} \label{serrs}
\mathcal{W}^{(2N)}_{ST}(x,\, k;\,\beta) = \sum_{\eta=0}^N\,\hbar^{2\eta}\,\left(\frac{\omega}{k_BT}\right)^{2\eta}\mathcal{W}_{2\eta}(x,\, k;\,\beta),
\end{equation}
as an approximation to the form of $\sum_{\ell}\exp(- E_{\ell}/k_BT) \mathcal{W}_{\ell}(x,\, k)$, up to order $\mathcal{O}(\hbar^{2N})$, where quantum fluctuations are driven by $\hbar^{2\eta}$ coefficients.
This corresponds to the equilibrium equation devised by Wigner \cite{Wigner}, where the TD equilibrium conditions drive the quantum system to yield adiabatically the Wigner stationary distribution, $\mathcal{W}_{ST}(x,\, k;\,\beta)$, as $\lim_{N\to \infty}\mathcal{W}^{(2N)}_{ST}(x,\, k;\,\beta)$.

A systematic procedure for accounting for the $\mathcal{O}(\hbar^{2\eta})$ contributions to the series expansion, Eq.~ \eqref{serrs}, has been developed \cite{Novo2022} in order to be applied to any Hamiltonian, $\mathcal{H}(x,\,k) = \mathcal{K}(k) + \mathcal{V}(x)$, constrained by the condition $\partial^2 \mathcal{H} / \partial x \, \partial k = 0$.
To illustrate the case where contributions from $\mathcal{O}(\hbar^{4})$ or higher are suppressed, one has $\mathcal{W}_{2}(x,\, k;\,\beta) = \chi_{(x,\, k;\,\beta)}\, \mathcal{W}_{0}(x,\, k;\,\beta)$, such that the procedure from Ref.~\cite{Novo2022} results in
\begin{eqnarray}\label{oideem}
\mathcal{W}^{(2)}_{ST}(x,\,k;\,\beta) &=& \frac{\mathcal{Z}_0(\beta)}{ \mathcal{Z}_{ST}(\beta)}\,\mathcal{W}_{0}(x,\,k;\,\beta)
\left[
1 +\chi_{(x,\,k;\,\beta)}\right],
\end{eqnarray}
with
\begin{eqnarray}
\chi_{(x,\,k;\,\beta)} &=&
-\frac{\beta^2}{8}
\partial_x^{2}\mathcal{V}(x)\,\partial^2_k \mathcal{K}(k)
+\frac{\beta^3}{24}
\left[
\partial_x^{2}\mathcal{V}(x)\,\left(\partial_k \mathcal{K}(k)\right)^2
+
\partial_k^{2}\mathcal{K}(k)\,\left(\partial_x \mathcal{V}(x)\right)^2
\right],
\end{eqnarray}
where the perturbative contributions, considering any dependence of the involved functions on $\beta$ (cf. Ref.~\cite{Novo2022}), are rescaled into order $\mathcal{O}\left(\beta^3\right)$ corrections, and the pre-factor, ${\mathcal{Z}_0(\beta)}/{ \mathcal{Z}_{ST}(\beta)}$, has been introduced to ensure unitarity properties.

The Wigner currents are then written as \cite{Novo2022}
\begin{equation}
\mathcal{J}^{(2)}_x(x,\, k;\,\beta) = + \left\{ \partial_k \mathcal{K}(k) \left(1 + \chi_{(x,\, k;\,\beta)}\right) 
- \frac{1}{24}\,\partial_k^3 \mathcal{K}(k)\,
\left[\beta^2 \left(\partial_x \mathcal{V}(x)\right)^2 -\beta \partial^2_x \mathcal{V}(x)\right]\right\}\mathcal{W}_0,
\label{z500BB}
\end{equation}
and
\begin{equation}
\mathcal{J}^{(2)}_k(x,\, k;\,\beta) = - \left\{ \partial_x \mathcal{V}(x) \left(1 + \chi_{(x,\, k;\,\beta)}\right) 
- \frac{1}{24}\,\partial_x^3 \mathcal{V}(x)\,
\left[\beta^2 \left(\partial_k \mathcal{K}(k)\right)^2 -\beta \partial^2_k \mathcal{K}(k)\right]\right\}\mathcal{W}_0,
\label{z500BB}
\end{equation}
where the non-Liouvillian $\mathcal{O}(\beta^2)$ feature is identified by the non-vanishing value of $\mbox{\boldmath $\nabla$}_{\xi} \cdot \mathbf{w}$ written as
\begin{equation}
\mbox{\boldmath $\nabla$}_{\xi} \cdot \mathbf{w} = \frac{\beta^2}{12}
\left(
\partial_k^3 \mathcal{K}(k)\partial^2_x \mathcal{V}(x)\partial_x \mathcal{V}(x)
-
\partial_x^3 \mathcal{V}(x)\partial^2_k \mathcal{K}(k)\partial_k \mathcal{K}(k)\right).
\end{equation}

Considering the prey-predator Hamiltonian, Eq.~\eqref{Ham}, for $\mathcal{K}(k) = k + e^{-k}$ and $\mathcal{V}(x)= a(x + e^{-x})$, one firstly writes the LV associated classical distribution as
\begin{equation}
\mathcal{W}_0(x,\,k;\,\, \beta) = \mathcal{Z}^{-1}_0(\beta)\,\exp\left\{-\, \beta \left[k + e^{-k} +a\left(x + e^{-x}\right)\right]\right\},
\end{equation}
with the phase-space domain $x \in (-\infty,\,+\infty)$ and $k \in (-\infty,\,+\infty)$ ($z \in [0,\,+\infty)$ and $y \in [0,\,+\infty$). These lead to 
\begin{equation}
\mathcal{Z}_0(\beta) = 
\int^{+\infty}_{-\infty} \hspace{-.35cm}dx\,\int^{+\infty}_{-\infty} \hspace{-.35cm}dk\,
\mathcal{W}_0(x,\,k;\,\, \beta)=
\frac{1}{\left(a^a\beta^{a+1}\right)^{\beta}} \, \Gamma(\beta)\, \Gamma(a\beta),
\end{equation} 	 
where $\Gamma[s]$ is Euler's gamma function \cite{Gradshteyn}.

After some straightforward manipulations, from Eq.~\eqref{oideem}, one obtains
\begin{eqnarray}\label{oideem2}
\mathcal{W}^{(2)}_{ST}(x,\,k;\,\, \beta) &=& \frac{\mathcal{Z}_0\, (\beta)}{ \mathcal{Z}_{ST}\, (\beta)}\,\mathcal{W}_{0}(x,\,k;\,\beta) \,
\left[1+\chi^{(LV)}(x,\,k;\,\beta)
\right],
\end{eqnarray}	
with
\begin{equation}
\chi^{(LV)}_{(x,\,k;\,\beta)} = 
\frac{a \beta^2}{8}\left[
\frac{4\beta}{3}
\left(
a\, \sinh^2(x/2) + \sinh^2(k/2)
\right)
- 1
\right]e^{-(k+x)},
\end{equation}
\begin{equation}\mathcal{Z}_{ST}(\beta) = 
\frac{1}{\left(a^a\beta^{a+1}\right)^{\beta}}\left(1-\frac{a\,\beta^2}{24}\right) \, \Gamma(\beta)\, \Gamma(a\beta),
\end{equation} 
from which the associated Wigner currents are given by
\begin{eqnarray}
\mathcal{J}^{(2)}_x(x, \, k;\,\beta) &=& \left[(1-e^{-k})\left(1+\chi^{(LV)}_{(x,\,k;\,\beta)}\right) + 
a\,\frac{\beta}{24}\left(
4a\,\beta \sinh^2(x/2) - 1
\right)e^{-(k+x)}
\right]
\mathcal{W}_{0},\\
\mathcal{J}^{(2)}_k(x, \, k;\,\beta) &=& -a\left[(1-e^{-x})\left(1+\chi^{(LV)}_{(x,\,k;\,\beta)}\right) +
\frac{\beta}{24}\left(
4\,\beta \sinh^2(k/2) - 1
\right)e^{-(k+x)}
\right]
\mathcal{W}_{0},
\end{eqnarray}	
and the Liouvillianity quantifier is given by
\begin{equation}\label{groms}
\mbox{\boldmath $\nabla$}_{\xi} \cdot \mathbf{w} = \frac{a\, \beta^2}{12}
\left(a\,e^{-x} -e^{-k}
\right)e^{-(k+x)}.
\end{equation}

\begin{figure}[h]
$(a)$\includegraphics[scale=0.33]{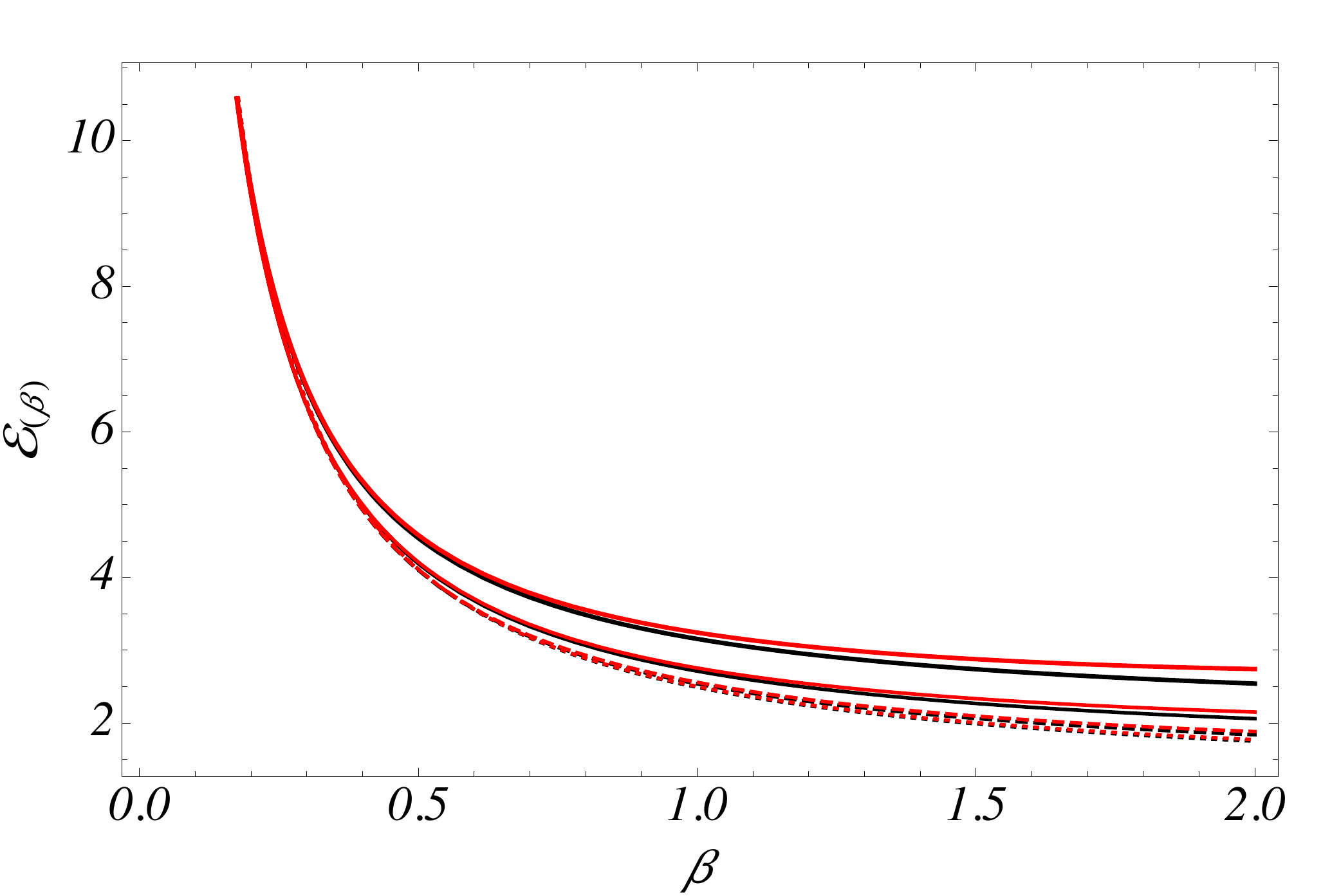}\hspace{-.1 cm}
$\qquad (b)$\includegraphics[scale=0.33]{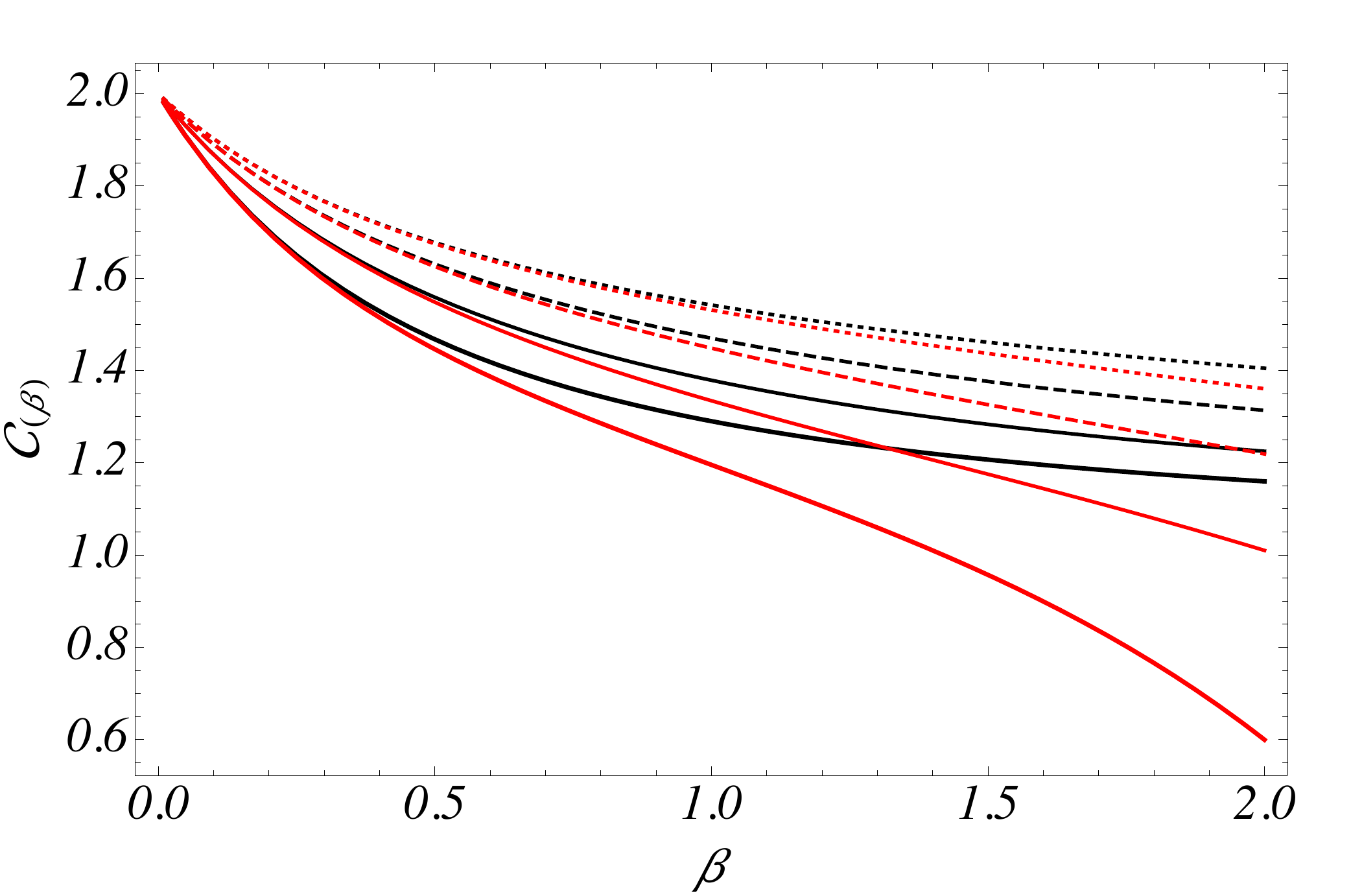}\
\renewcommand{\baselinestretch}{.6}
\caption{\footnotesize
(Color online) (a) Internal energy, $\mathcal{E}(\beta)$, and (b) heat capacity, $\mathcal{C}(\beta)$, as function of $\, \beta$, for classical (dark black lines) and quantum ( $\mathcal{O}(\hbar^2)$) (light red lines) stationary regimes.
Results are for $a = 1/2$ (dotted lines), $1$ (dashed lines), ${2}$ (thin lines) and $4$ (thick lines).}
\label{HarperHarper02}
\end{figure}

For microscopic systems as, for instance, in the context of molecular and biochemical systems \cite{PP00,PP01,PP02,PP03,PP04}, the above results for the partition functions for classical and quantum ($\mathcal{O}(\hbar^2)$) stationary ensembles, $\mathcal{Z}_{ST}(\beta)$ and $\mathcal{Z}_{0}(\beta)$, can be employed for obtaining the associated (dimensionless) internal energy, $\mathcal{E}_{\, \beta} (\equiv E/(\hbar\omega))$, and heat capacity, $\mathcal{C}_{\, \beta}(\equiv C/k_B)$, given respectively by
$$
\mathcal{E}_{\, \beta} = -\frac{\partial~}{\partial \, \beta}\ln\left[{\mathcal{Z}(\beta)}\right]
,\quad\mbox{and} \quad
 \mathcal{C}_{\, \beta} = \beta^2\left(\frac{\partial~}{\partial \, \beta}\right)^2\ln\left[{\mathcal{Z}(\beta)}\right],$$
whose behavior are depicted in Fig.~\ref{HarperHarper02}.

\begin{figure}
\hspace{-.8 cm}\includegraphics[scale=0.27]{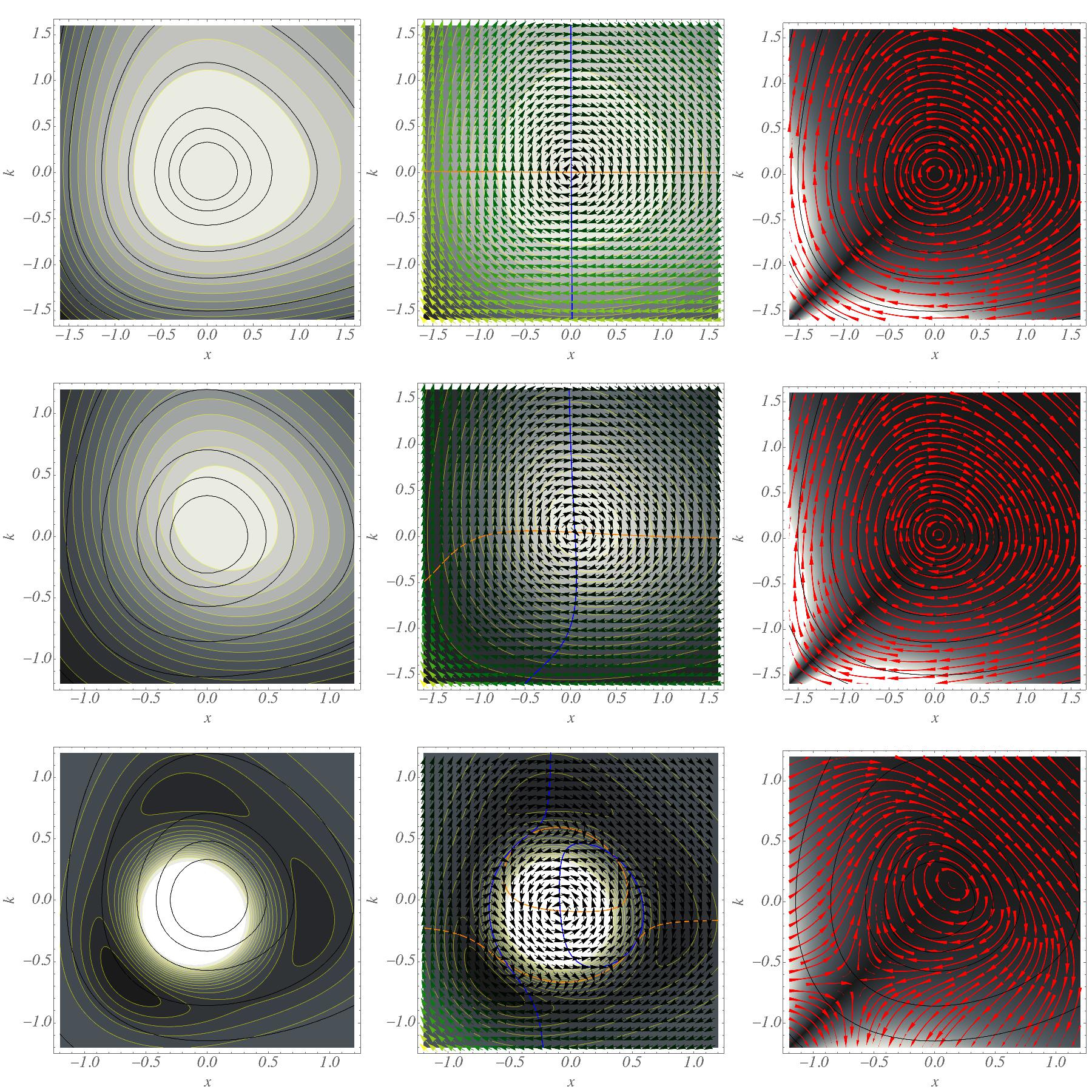}
\renewcommand{\baselinestretch}{.6}
\caption{\footnotesize
(Color online)
{\it First column}: Classical (dark black contours) and QM corrected (light yellow contours) Wigner function profiles, $\mathcal{W}_0(x,\,k;\, \beta)$ and $\mathcal{W}^{(2)}_{ST}(x,\,k;\, \beta)$. The background depicts the quantum modified Wigner function, $\mathcal{W}^{(2)}_{ST}(x,\,k;\, \beta)$.
{\it Second column}: Features of the Wigner flow for the TD ensemble in the $x - k$ plane. Dark blue contour lines are for $J^{(2)}_x(x,\,k;\, \beta) = 0$ and light orange contour lines are for $J^{(2)}_k(x,\,k;\, \beta) = 0$.
The {\it green-yellow (dark-light)} vector arrow color scheme shows the (quantum $\mathcal{O}(\hbar^2)$) Wigner currents, $\mbox{\boldmath $\mathcal{J}$}^{(2)}(x,\,k;\, \beta)$ with the domains of quantum fluctuations bound by blue (dark) and orange (light) lines. Again, the background depicts the quantum modified Wigner function, $\mathcal{W}^{(2)}_{ST}(x,\,k;\, \beta)$.
{\it Third column}: Liouvillian quantifier (in the background), $\mbox{\boldmath $\nabla$}_{\xi} \cdot \mathbf{w}$, superposed by the normalized quantum velocity field representation (red arrows), $\mathbf{w} /\vert\mathbf{w}\vert$. The background color scheme (from darker, $\mbox{\boldmath $\nabla$}_{\xi} \cdot \mathbf{w} \sim 0$, to lighter regions, $\mbox{\boldmath $\nabla$}_{\xi} \cdot \mathbf{w} > 0$) reinforces the approximated Liouvillian behavior for the axis $x=k$.
The results are for $a=1$ (isotropic $x-k$ distribution) and $\beta =0.1$ (first row), $1$ (second row) and $5$ (third row), from which one notices that the quantum distortions become relevant only for lower temperatures (increasing $\, \beta$ values).}
\label{HarperHarper03}
\end{figure}

More relevantly, the non-Liouvillian behavior is depicted in Fig.~\ref{HarperHarper03}, and can be compared with the classical regime, which is parameterized by the collection of black thin lines. 
To pedagogically interpret Fig.~\ref{HarperHarper03}, for $\beta \ll 1$ (first row), quantum fluctuations are suppressed by the thermal fluctuations and cannot be identified through the pattern of the Wigner function (first collumn) and of the corresponding Wigner flow (second collumn).
The Liouvillian quantifier (third collumn) is the only tool that allows for visualizing the quantumness.
The obvious reason for this resides on the form of Eq.~\eqref{groms} where there are no terms to compete with the fluctuations proportional to $\beta^2$, as depicted by the dark-light background color scheme.
Of course, as mentioned above, such a quantum distorting map is sensible up to $\mathcal{O}(\hbar^2)$ features which are mostly evinced for $\beta > 1$. 
In this case, by confronting the behavior of the Wigner current (second column), $\mbox{\boldmath $\mathcal{J}$}$, with the quantum velocity (third column), $\mathbf{w}$, the effect of the Wigner function negative values given by $\mathcal{W}^{(2)}_{ST}(x,\,k;\, \beta)$, for the case where $\beta > 1$ (third row), are shown in blue (dark) and orange (light) contour lines, which correspond to $\mathcal{J}^{(2)}_k = 0$ and $\mathcal{J}^{(2)}_x = 0$, respectively. Notice that the normalized quantum velocity field representation (red arrows) are not affected by such a flux reversibility, which shows the complementary information brought up by $\mbox{\boldmath $\mathcal{J}$}$ and $\mathbf{w}$.
In such a context, as the temperature decreases, the Wigner flow stagnation points (second collumn) are identified by orange-blue (light-dark) crossing lines, namely for $\mathcal{J}^{(2)}_x = \mathcal{J}^{(2)}_k=0$.
For these effects to make sense as quantum fluctuations, for an arbitrarily chosen $\beta=5$, one should have $5 k_B T/\omega (= \hbar) \ll 1$ in order to ensure the validity of the $\mathcal{O}(\hbar^2)$ approximation.

As it will be discussed in the following section, gaussian ensembles are much more sensitive to higher order quantum corrections, given that, through their analytical properties, the contributions due to the $\hbar^2$ series expansion provide exact expressions for the Wigner currents.

\section{Gaussian ensembles}

Gaussian ensembles are introduced through a dimensionless distribution written as\footnote{The correspondence with the physical position and momentum variables, $q$ and $p$, is given through
\begin{equation}
G_\alpha(q,\,p) = \frac{\alpha^2}{\pi\hbar}\, \exp\left[-\frac{\alpha^2}{\hbar}\left(\frac{q^2}{\mathcal{A}^2}+ \mathcal{A}^2\,p^2\right)\right],
\end{equation}
so that $\mathcal{G}_\alpha(x,\,k) = \hbar \,G_\alpha(q,\,p)$ where $\mathcal{A}$ is identified by $\mathcal{A} = (m\,\omega)^{-1}$, for a mass scale, $m$, and an arbitrary angular frequency, $\omega$.}
\begin{equation}
\mathcal{G}_\alpha(x,\,k) = \frac{\alpha^2}{\pi}\, \exp\left[-\alpha^2\left(x^2+ k^2\right)\right],
\end{equation}
which replaces the Wigner function into Eqs.~\eqref{imWA} and \eqref{imWB}, leading to the following associated gaussian flow contributions,
\begin{eqnarray}
\label{imWA22}\partial_x\mathcal{J}^{\alpha}_x(x, \, k) &=& +\sum_{\eta=0}^{\infty} \left(\frac{i}{2}\right)^{2\eta}\frac{1}{(2\eta+1)!} \, \left[\partial_k^{2\eta+1}\mathcal{K}(k)\right]\,\partial_x^{2\eta+1}\mathcal{G}_{\alpha}(x, \, k),
\\
\label{imWB22}\partial_k\mathcal{J}^{\alpha}_k(x, \, k) &=& -\sum_{\eta=0}^{\infty} \left(\frac{i}{2}\right)^{2\eta}\frac{1}{(2\eta+1)!} \, \left[\partial_x^{2\eta+1}\mathcal{V}(x)\right]\,\partial_k^{2\eta+1}\mathcal{G}_{\alpha}(x, \, k).
\end{eqnarray}
The remarkable property that can be derived from the above expressions \cite{Novo2022} are retrived by means of the gaussian derivatives,
\begin{equation}\label{ssae}
\partial_\upsilon^{2\eta+1}\mathcal{G}_{\alpha}(x, \, k) = (-1)^{2\eta+1}\alpha^{2\eta+1}\,\mbox{\sc{h}}_{2\eta+1} (\alpha \upsilon)\, \mathcal{G}_{\alpha}(x, \, k),\qquad\mbox{for $\upsilon = x,\, k$,}
\end{equation}
where $\mbox{\sc{h}}_n$ are the Hermite polynomials of order $n$, 
through which the complete ($\eta \to \infty$) series expansions, Eqs.~\eqref{imWA22} and \eqref{imWB22}, can be converted into convergent functions that account for the quantum distortions on the classical pattern \cite{Novo2022}.

For the prey-predator system dynamics, Eq.~\eqref{Ham}, where $\mathcal{K}(k) = k + e^{-k}$ and $\mathcal{V}(x)= a(x + e^{-x})$, the derivatives give
\begin{eqnarray}
\label{t111B}
\partial_x^{2\eta+1}\mathcal{K}(k) &=& \delta_{\eta 0} - e^{-k},\\
\label{t222B}
\partial_k^{2\eta+1}\mathcal{V}(x) &=& a \left(\delta_{\eta 0} - e^{-x}\right), 
\end{eqnarray}
which, once inserted into Eqs.~\eqref{imWA22} and \eqref{imWB22}, after taking into account the Hermite polynomials property,
\begin{equation}
\sum_{\eta=0}^{\infty}\mbox{\sc{h}}_{2\eta+1} (\alpha \upsilon)\frac{s^{2\eta+1}}{(2\eta+1)!} = \sinh(2s\,\alpha\upsilon) \exp[-s^2],
\end{equation}
lead, according to the procedure from Ref.~\cite{Novo2022}, to
\begin{eqnarray}
\label{imWA4CC3}
\partial_x\mathcal{J}^{\alpha}_x(x, \, k) &=& -2 \left[\alpha^2 \,x - 
\sin\left(\alpha^2\,x\right)\,e^{\frac{\alpha^2}{4}-k}
\right]
\mathcal{G}_{\alpha}(x, \, k),\\
\label{imWB4CC3}
\partial_k\mathcal{J}^{\alpha}_k(x, \, k) &=& +2a\left[\alpha^2\,k - 
\sin\left(\alpha^2\,k\right)\,e^{\frac{\alpha^2}{4}-x}
\right]
\mathcal{G}_{\alpha}(x, \, k),
\end{eqnarray}
where the exact dependence on Planck's constant, $\hbar$, is implicit from the assignments $x = \left(m\,\omega\,\hbar^{-1}\right)^{1/2} q$ and $k = \left(m\,\omega\,\hbar\right)^{-1/2}p$.
It shows that gaussian ensembles driven by the LV Hamiltonian allow for an analytical quantification of quantum fluctuations over the classical regime. Of course, the stationarity quantifier, $\mbox{\boldmath $\nabla$}_{\xi}\cdot \mbox{\boldmath $\mathcal{J}^{\alpha}$}$, obtained from Eqs.~\eqref{imWA4CC3} and \eqref{imWB4CC3}, is constrained by the Hamiltonian, Eq.~\eqref{Ham}.
Anyway, the Liouvillian quantifier, $\mbox{\boldmath $\nabla$}_{\xi} \cdot \mathbf{w}$, exhibits the complete pattern of the associated Wigner flow.

Through simple position and momentum integrations, over Eqs.~\eqref{imWA4CC3} and \eqref{imWB4CC3},
the corresponding Wigner currents can be expressed in terms of error functions, $\mbox{\sc{Erf}}[\dots]$,
\begin{eqnarray}
\label{imWA4CCD4}\mathcal{J}^{\alpha}_x(x, \, k) &=& 
\mathcal{G}_{\alpha}(x, \, k)
-\frac{i\,\alpha}{2\sqrt{\pi}} \,e^{-(k+\alpha^2 k^2)}
\left\{\mbox{\sc{Erf}}\left[\alpha(x-i/2)\right]-\mbox{\sc{Erf}}\left[\alpha(x+i/2)\right]\right\},\\
\label{imWB4CCD4}\mathcal{J}^{\alpha}_k(x, \, k) &=& 
-a\,\mathcal{G}_{\alpha}(x, \, k)
+\frac{i\,a\,\alpha}{2\sqrt{\pi}} \,e^{-(x+\alpha^2 x^2)}
\left\{\mbox{\sc{Erf}}\left[\alpha(k-i/2)\right]-\mbox{\sc{Erf}}\left[\alpha(k+i/2)\right]\right\}.\,\,\,\,
\end{eqnarray}

The above results show how gaussian ensembles are much more sensitive to higher order quantum corrections, given that from the results of Eqs.~\eqref{ssae}-\eqref{imWB4CCD4}, the contributions due to the $\hbar^2$ series expansion provide exact expressions for the Wigner currents. Of course, for simpler Hamiltonian systems the same strategy applies. For the quartic oscillator, for instance, the series Eqs.~\eqref{imWA22}-\eqref{imWB22} are automatically truncated at order 2 and 4 due to $k^2$ and $x^4$ contributions, respectively. In this case, considering gaussian ensembles is not relevant once that the Hamiltonian, even if driving a non-linear dynamical system, has a finite polynomial form which truncates Eqs.~\eqref{imWA22}-\eqref{imWB22} at the maximal power of the respective polynomial. For analytical functions as $\mathcal{K}(k) = k + e^{-k}$ and $\mathcal{V}(x) = a(x + e^{-x})$, the series expansion could only be forcedly truncated, resulting in an approximated solution which would inaccurate in accounting for quantum distortions coupled to non-linear effects. Gaussian ensembles, through manipulations which follows from Eq.~\eqref{ssae}, avoid this issue since the resulting series is convergent.

\begin{figure}[h!]
\vspace{-1.4 cm}\includegraphics[scale=0.2]{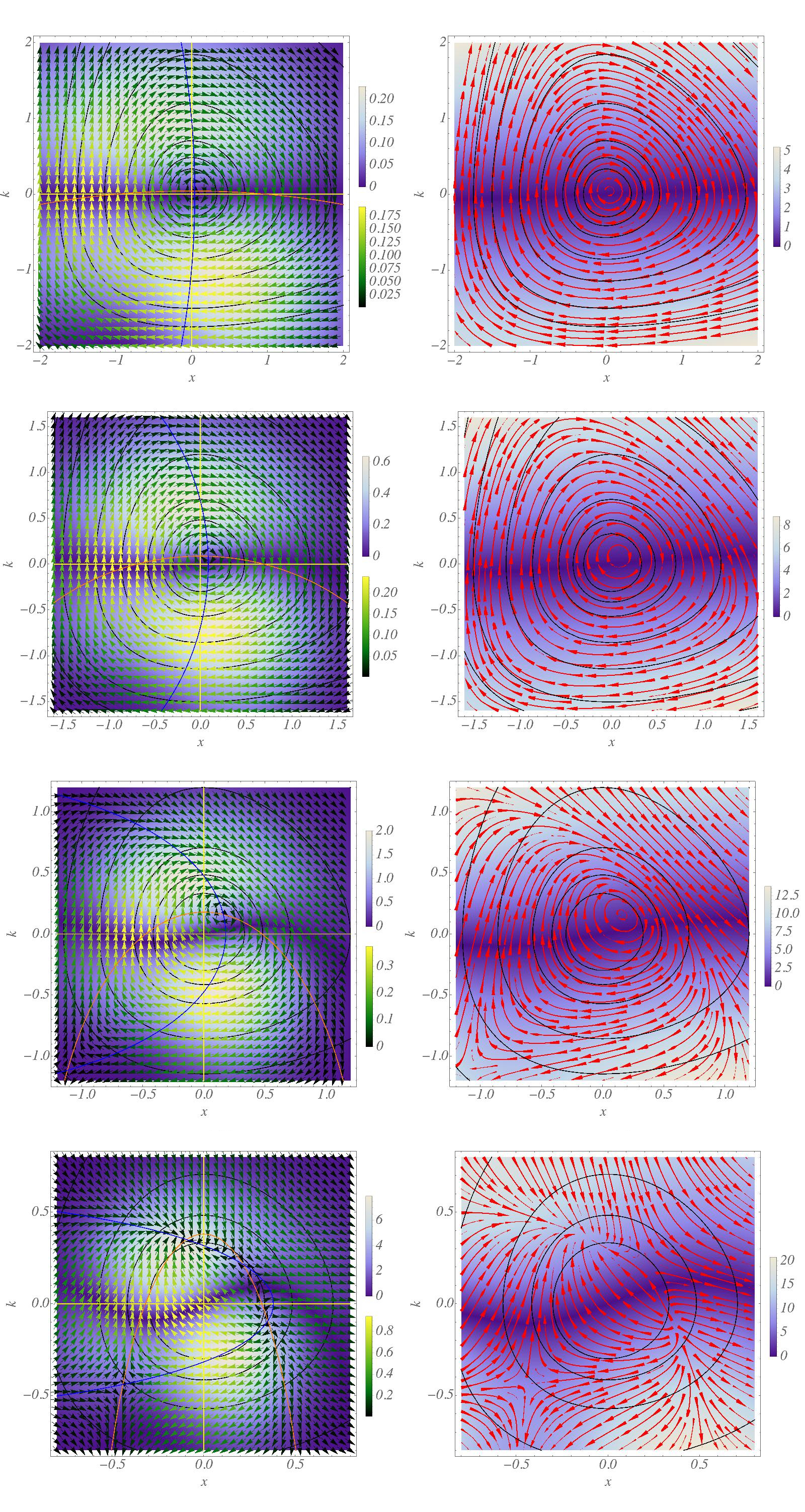}
\renewcommand{\baselinestretch}{.6}
\vspace{-.6 cm}\caption{\footnotesize
(Color online)
{\it First column}: Wigner currents (vector scheme), $\mbox{\boldmath $\mathcal{J}^{\alpha}$}$, and stationarity quantifier (background scheme), $\mbox{\boldmath $\nabla$}_{\xi} \cdot \mbox{\boldmath $\mathcal{J}^{\alpha}$}$, for the gaussian ensemble in the $x - k$ plane. At $\tau = 0$, gaussian ensembles do not exhibit neither vortices nor stagnation points, in a kind of camouflage of the quantum distortions. The magnitude of $\mbox{\boldmath $\mathcal{J}^{\alpha}$}$ is given by the {\it green-yellow (dark-light)} color scheme.
The stationarity quantifier, $\mbox{\boldmath $\nabla$}_{\xi} \cdot \mbox{\boldmath $\mathcal{J}$}$, and the modulus of $\mbox{\boldmath $\mathcal{J}$}$ are described according to the {\it dark-light blue background} and {\it blue-green-yellow (dark-light) vector arrow} color schemes, respectively. The results are for the increasing spreading pattern of the gaussian function, from $\alpha =1/\sqrt{2}$ (first row), $1$ (second row) and $\sqrt{2}$ (third row). Peaked gaussians ($\alpha =\sqrt{2}$) localize the quantum distortions which result in non-stationarity.
{\it Second column}: Normalized quantum velocities (red arrows), $ \mathbf{w}$, and the Liouvillian quantifier, $\mbox{\boldmath $\nabla$}_{\xi} \cdot \mathbf{w}$, depicted through the background color scheme, from darker, $\mbox{\boldmath $\nabla$}_{\xi} \cdot \mathbf{w} \sim 0$, to lighter regions, $\mbox{\boldmath $\nabla$}_{\xi} \cdot \mathbf{w} > 0$. 
The classical pattern is shown as a collection of black lines.}
\label{Lotka04}
\end{figure}

The constraints over the Wigner currents from Eqs.~\eqref{imWA4CCD4}-\eqref{imWB4CCD4} arise from $\mbox{\boldmath $\mathcal{J}^{\alpha}$} = \mathbf{w}\,\mathcal{G}_{\alpha}(x, \, k)$, providing the expressions for the components of the quantum velocity, $\mathbf{w}$, $w_x$ and $w_k$, and the topological properties of the Wigner flow.
The quantum distortions show up on the Wigner current pattern through separatrix intersections and saddle points (winding number equals to $0$), as well as by clockwise and anti-clockwise vortices (winding number equals to $+1$ and $-1$), as depicted in the first column of Fig.~\ref{Lotka04}.
The Wigner continuity equation, Eq.~\eqref{z51}, imposes the emergence of quantum contra-flux fringes to compensate the evolution effects due to the the quantum vortices and saddle points, while the classical LV profile does not show this behavior.
As before, blue (dark) and orange (light) contour lines corresponding to $\mathcal{J}^{\alpha}_k = 0$ and $\mathcal{J}^{\alpha}_x = 0$, respectively, illustrates this emergence of quantum features, which, as depicted in Fig.~\ref{Lotka04}, clearly change the coordinates of the equilibrium points (i.e. the phase-space attractors).

The stationarity quantifier, $\mbox{\boldmath $\nabla$}_{\xi} \cdot \mbox{\boldmath $\mathcal{J}^{\alpha}$}$, is described by the background color scheme, so that lighter regions correspond to non-vanishing local contributions to $\partial_{\tau} \mathcal{G}_{\alpha}(x,\,k)$. In spite of Wigner currents and quantum velocities being constrained by $\mbox{\boldmath $\mathcal{J}^{\alpha}$} = \mathbf{w}\,\mathcal{G}_{\alpha}$, stationarity and Liouvillianity are smoothly decoupled from each other, due to the gaussian spreading effect.
The onset configuration ($\tau = 0$) of the gaussian Wigner flow is depicted in Fig.~\ref{Lotka04}, where the density plot shows also the stationarity and Liouvillianity quantifieres, $\mbox{\boldmath $\nabla$}_{\xi} \cdot \mbox{\boldmath $\mathcal{J}^{\alpha}$}$ and $\mbox{\boldmath $\nabla$}_{\xi} \cdot \mathbf{w}$.

\begin{figure}
\includegraphics[scale=0.47]{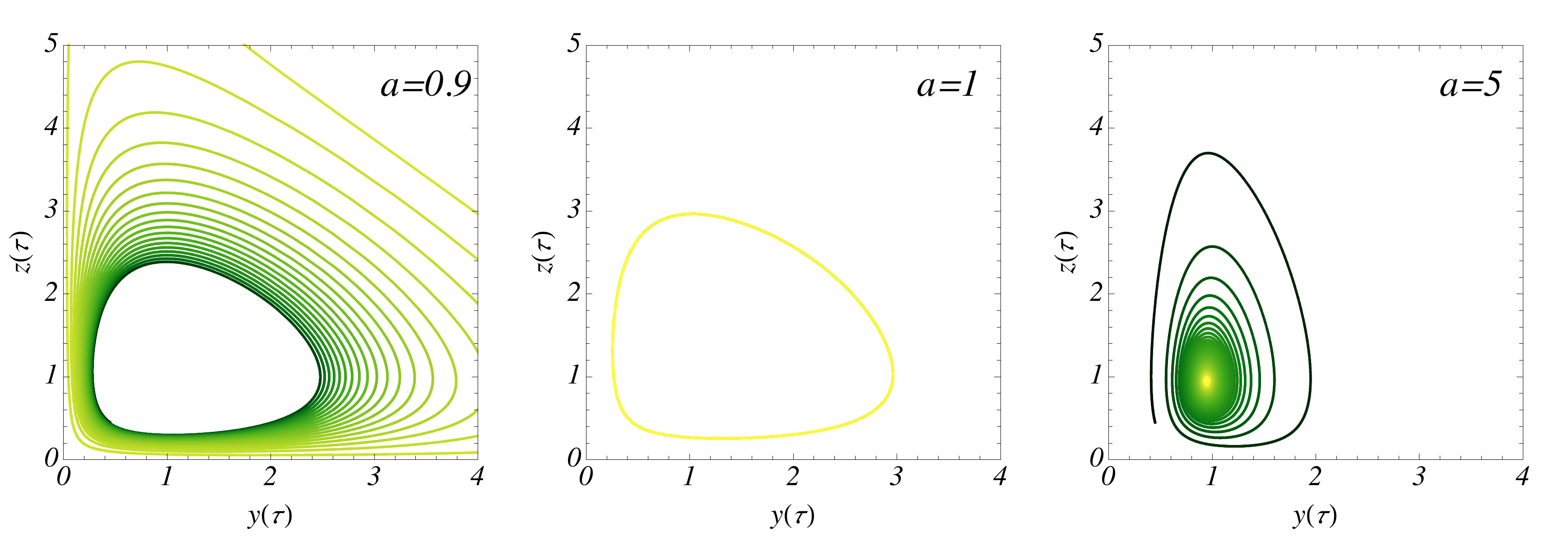}
\renewcommand{\baselinestretch}{.6}
\caption{\label{NovaNova}\footnotesize
(Color online)
Phase space oscillating population portrait impacted by quantum corrections modulated by gaussian ensembles. The green-yellow (dark-light) color scheme describes the phase-space trajectory evolution, $(y(\tau),\,z(\tau))$, from $\tau = 0$ (dark green tone) to $\tau \gg 0$ (light yellow tone). Plots are for $a=0.9,\, 1$ and $5$, and $\alpha = 0.8$.
.}
\end{figure}

The impact of quantum corrections can also be evaluated in terms of the semi-classical analysis of the so-called hyperbolic equilibrium and stability conditions described in terms of $y(\tau)$ and $z(\tau)$ variables. By expanding Eqs.~\eqref{imWA4CCD4} and \eqref{imWB4CCD4} up to order $\mathcal{O}(\alpha^4)$, and recasting the associate quantum-analogue velocities in terms of $y(\tau)$ and $z(\tau)$ time derivatives, one has
\begin{eqnarray}
d{y}/d\tau &\approx& z\, y\left\{1+\frac{\alpha^2}{12} + \frac{\alpha^4}{160}[1-\ln(y^{\frac{1}{3}})]\right\}-y,\nonumber\\
d{z}/d\tau &\approx& a z -a z\, y\left\{1+\frac{\alpha^2}{12} + \frac{\alpha^4}{160}[1-\ln(y^{\frac{1}{3}})]\right\},
\end{eqnarray}
From such effective quantum modified LV equations, further analysis of stability of the equilibrium can be evaluated in terms of the Jacobian matrix identified by
\begin{equation}
j (y,\,z) = \left[\begin{array}{rr}
\partial_y f(y,\,z) & \partial_z f(y,\,z)\\
\partial_y g(y,\,z) & \partial_z g(y,\,z)
\end{array}\right],\nonumber
\end{equation}
so that they can be stratified into subclassifications, through the trace, ${Tr}[\dots]$, and the determinant, ${Det}[\dots]$, of $j (y,\,z)$, when all derivatives are evaluated at the equilibrium points obtained from $f(y_e,\,z_e)=g(y_e,\,z_e)=0$.
Second-order contributions, from $\alpha^2$, just change the equilibrium points from $y_e=z_e=1$ (i.e. $z_e=y_e=0$) to $y_e=z_e=(1+\alpha^2/12)^{-1}$ and do not affect the stability condition of closed orbits in the phase-space, which are identified by $Tr[j(y_e,\,z_e)] =0$ for $Det[j(y_e,\,z_e)] > 0$.
According to the hyperbolic stability criterium, focus and node stabilities are defined by trace properties as 
\begin{eqnarray}
Tr[j(y_e,\,z_e)] > 0 \qquad &\to& \mbox{instability},\nonumber\\
Tr[j(y_e,\,z_e)] < 0 \qquad &\to& \mbox{stability},\nonumber
\end{eqnarray}
when $Det[j(y_e,\,z_e)] > 0$\footnote{Being not relevant in this case, the threshold for focus and nodes is identified by $\Delta[j] = Tr[j]^2 - 4 Det[j] = 0$ such that
$\Delta[j(y_e,\,z_e)] > 0$, for nodes, and $\Delta[j(y_e,\,z_e)] < 0$ for focus.}. Saddle points occur for $Det[j(y_e,\,z_e)] < 0$.
Hence, considering an iterative contribution of $\mathcal{O}(\alpha^4)$, one has
\begin{eqnarray}
Tr[j(y_e,\,z_e)] \approx (1-a)\frac{\alpha^4}{160}\qquad \mbox{and}\qquad Det[j(y_e,\,z_e)]\approx a\left(1 + \frac{\alpha^4}{80}\right),\nonumber
\end{eqnarray} 
from which one notices that $\mathcal{O}(\alpha^4)$ corrections drive the system to stable ($a > 1$) and unstable ($a < 1$) domains as depicted in Fig.~\ref{NovaNova}.
Plots from Fig.~\ref{NovaNova} are obtained by numerical solving the exact expressions for time derivative of $y(\tau)$ and $z(\tau)$, obtained from Eqs.~\eqref{imWA4CCD4} and \eqref{imWB4CCD4}.
Interpreting such phase-space oscillating population portraits as the {\it quantum analog} of a prey-predator-like dynamics (cf. Hamiltonian Eq.~\eqref{Ham}), allows for stating that for $a < 1$ and $\alpha \neq 0$, the quantum effects asymptotically drive the system to periodic population extinctions and revivals. 
Likewise, for $a > 1$ and $\alpha \neq 0$, the population oscillations are asymptotically suppressed, i.e. they approximately stabilize around the modified equilibrium points $y_e=z_e=(1+\alpha^2/12)^{-1}$. 

Therefore, one can notice that the {\it quantum analog} hypothesis of the LV dynamics discussed here, for which either $[x,\,k] = i$, or simply $[x,\,k] \neq 0$, results in quantum effects which, for gaussian ensembles, lead to an explicit reformulation of the equations of motion parameterized by $\mathbf{w} = \mbox{\boldmath $\mathcal{J}^{\alpha}$} / \mathcal{G}_{\alpha}$, which brings up novel elements for the stability analysis of prey-predator dynamics.
Once that the phenomenological behavior can be examined and expressed in terms of the evolution of (gaussian) ensembles of species $y = e^{-x}$ and $z = e^{-k}$, results for stable stationary {\it quantum analog} of classical (Liouvillian) patterns can be macroscopically identified and related to non-commutative features and ensemble parameters.

\subsection{Stationary gaussian ensembles -- a quantum camouflage}

The list of quantum-mechanical systems with eigenstate analytical solutions is constrained by the existence and uniqueness of the solutions of the linear partial differential associated problem, i.e., the Schr\"{o}dinger equation. 
For one-dimensional systems, the problem can be read as a second order linear ordinary differential equation driven by the so-called quadratic momentum operator, $\mathcal{K}(k)=k^2\equiv -\partial_x^2$, for which existence and uniqueness conditions obviously apply. Even for the reduced one-dimensional scenario, the vast majority of classical analogous problems, for which the Hamilton's equations of motion can be analytically solved, has not a Schr\"{o}dinger quantum mechanical counterpart which admits analytical solutions. Of course, considering perturbation theory, approximative methods are highly efficient in circumvent such limitations. However, more constrained integrable classical systems, once driven by Hamiltonians with non-quadratic contributions from momentum dependent operators, do not admit being cast in the Schr\"{o}dinger equation usual linear differential form, given that $\mathcal{K}(k)\neq k^2$. The LV system discussed here, for which $\mathcal{K}(k) = k + e^{-k}$, as well as the Harper system discussed in Ref. \cite{Novo2022}, for which $\mathcal{K}(k) = \cos(k)$, are typical examples of non-linear integrable dynamical systems with classical solutions for which the quantum analogue problem does not fit the Schr\"{o}dinger equation in its linear form.

Fortunately, the Schrödinger equation is not the only way to investigate quantum mechanical systems and to make predictions. The generalized phase-space description of non-linear Hamiltonian systems discussed above \cite{Novo2022} joins the phase-space frameworks addressed in the previous section \cite{Hillery,Zurek01,Zurek02,Berry79,DeAlmeida98,Hannay82} for investigating classical-quantum correspondence of systems which does not rely on the Schr\"odinger equation in its linear form.

Given the mathematical nature of the exact Wigner currents involving the gaussian ensembles in the above context, one could raise the question if such a scheme could be helpful for identifying analytical solutions for the Hamiltonian eigenstates in a problem corresponding to a Schr\"odinger-like equation with the same non-linear correspondence to the momentum operator.
The answer is positive for a very particular form of the Hamiltonian ${\mathcal{H}}(x,\,k)$, now re-written as ${\mathcal{H}}(x,\,k) = \tilde{\mathcal{K}}(k) + \tilde{\mathcal{V}}(x)$, with
\begin{eqnarray}
\tilde{\mathcal{V}}(x) &=& \cosh(\nu_1\,x) + \lambda_x \,\cos(\nu_2\,x),\nonumber\\
\label{seila2}\tilde{\mathcal{K}}(k) &=& \cosh(\mu_1\,k) + \lambda_k \,\cos(\mu_2\,k),\nonumber
\end{eqnarray}
with arbitrary $\lambda_k$ and $\lambda_x$, whose phase-space closed classical orbits are similar to those describing the LV system.
One can verify that, besides the classical stationarity for TD ensembles trivially derived from the $\partial^2 \mathcal{H} / \partial x \, \partial k = 0$ condition (cf. Eqs.~\eqref{z500BB2} and \eqref{z500CC2} inserted into Eq.~\eqref{z51dim}, with $\mathcal{W}_0$ from Eq.~\eqref{TDclass}), quantum gaussian ensembles, $\mathcal{G}_{\alpha}(x,\,k;\,\tau)$, also drive the quantum Wigner flow into the stationary regime. 
Given that the Schr\"odinger analog problem for an eigenfunction $\psi(x)$ would require the resolution of an infinite order differential equation,
\begin{eqnarray}
\label{seila1m}\cosh(i\,\mu_1\,\partial_x) + \lambda_k \,\cos(i\,\mu_2\,\partial_x)+\cosh(\nu_1\,x) + \lambda_x \,\cos(\nu_2\,x)]\psi(x)&=&0\nonumber\\
\Rightarrow \left\{\sum_{s=0}^{\infty}\left[(-1)^s\mu_1^{2s} + \lambda_k \mu_2^{2s}\right]\frac{\partial_x^{2s}}{(2s)!}+\cosh(\nu_1\,x) + \lambda_x \,\cos(\nu_2\,x)\right\}\psi(x)&=&0,\nonumber
\end{eqnarray}
from which the non-linear aspects are evinced. In principle, an analytical solution for $\psi(x)$ would not be realized without the results obtained from the manipulation of the infinite series expansion obtained for gaussian ensembles.

To verify the solution hypothesis, one uses Eqs.~\eqref{imWA22}-\eqref{ssae} for $\tilde{\mathcal{K}}(k)$ and $\tilde{\mathcal{V}}(x)$, and
\begin{eqnarray}
\label{t111BC}
\partial_x^{2\eta+1}\tilde{\mathcal{V}}(x) &=& \nu_1^{2\eta+1}\,\sinh(\nu_1\,x) - (-1)^{\eta}\,\lambda_x\,\nu_2^{2\eta+1}\,\sin(\nu_2\,x),\\
\label{t222BC}
\partial_k^{2\eta+1}\tilde{\mathcal{K}}(k) &=&\mu_1^{2\eta+1}\sinh(\mu_1\,k) - (-1)^{\eta}\,\lambda_k\,\mu_2^{2\eta+1}\, \sin(\mu_2\,k),
\end{eqnarray}
from which, after some straightforward mathematical manipulations, one obtains
\begin{eqnarray}
\label{imWA22mm}\partial_x\mathcal{J}^{\alpha}_x &=& +2\left[
\lambda_k\,\sin(\mu_2\,k)\,\sinh(\alpha^2\mu_2\,x)\,e^{-\frac{\alpha^2\mu^2_2}{4}} - \sinh(\mu_1\,k)\,\sin(\alpha^2\mu_1\,x)\,e^{+\frac{\alpha^2\mu^2_1}{4}}
\right]
\mathcal{G}_{\alpha}(x, \, k),\,\,\,\,\quad
\\
\label{imWB22mm}\partial_k\mathcal{J}^{\alpha}_k &=& -2\left[
\lambda_x\,\sin(\nu_2\,x)\,\sinh(\alpha^2\nu_2\,k)\,e^{-\frac{\alpha^2\nu^2_2}{4}} - \sinh(\nu_1\,x)\,\sin(\alpha^2\nu_1\,k)\,e^{+\frac{\alpha^2\nu^2_1}{4}}
\right]
\mathcal{G}_{\alpha}(x, \, k).\,\,\,\,\quad
\end{eqnarray}
It can be noticed that, from the condition $\alpha^2 =1$, with $\mu_{1(2)} = \nu_{2(1)}$, and $\lambda_{k(x)} = -\exp\left[{\nu^2_{2(1)}/{2}}\right]$, one has $\mbox{\boldmath $\nabla$}_{\xi} \cdot \mbox{\boldmath $\mathcal{J}^{\alpha}$} = \partial_x\mathcal{J}^{\alpha}_x+\partial_k\mathcal{J}^{\alpha}_k=0$.

In this case, the stationary behavior, which does not follow a Liouvillian pattern (i.e. $\mbox{\boldmath $\nabla$}_{\xi} \cdot \mathbf{w}\neq 0$) can be interpreted as a semi-classical {\it camouflage} of quantum distortions.
Etymologically, the concept of {\it camouflage} refers to a concealment of some object properties by the combination of methods that are unnoticed. 
On one hand, one has the stationary behavior of the classical TD ensemble of a non-linear system emulated by a gaussian ensemble which evolves quantum mechanically as a non-Liouvillian stationary state.
On the other hand, one has a non-linear classical Hamiltonian dynamics for which the quantum analog reproduces the ground state of the one-dimensional harmonic oscillator, which exhibits a Liouvillian pattern.

Naturally, it can be shown, even through the Schr\"odinger-like Eq.~\eqref{seila1m}, that the gaussian ensemble, $\mathcal{G}_{1}(x, \, k)$, is the zero-mode of $\tilde{\mathcal{H}}(x, \, k)$, i.e. $\tilde{\mathcal{H}}\,\mathcal{G}_{1} = 0$\footnote{A result that can also be extended to squeezed gaussian ensembles, $\mathcal{G}_{1}(x\,e^{+\zeta}, \, k\,e^{-\zeta})$, for $\mu_{1(2)} = e^{-2\zeta} \nu_{2(1)}$ and $\lambda_{k(x)} = -\exp\left[{e^{-2\zeta}\,\nu^2_{2(1)}/{2}}\right]$.}.

This particularity is not due to the form of Eqs.~\eqref{seila2}-\eqref{seila1m}, but arises from the property of quantum gaussian ensembles yielding well-behaved quantum currents for the non-linear Hamiltonian systems. That is, a kind of quantum {\it camouflage} rendered by the stationary behavior, a feature which deserves further investigation.

\section{Conclusions}

The phase-space WW framework for investigating the classical to quantum transition of systems driven by non-linear Hamiltonians, $\mathcal{H}(x,\,k)$, subjected to the condition $\partial^2 \mathcal{H} / \partial x \, \partial k = 0$, was discussed and analytically implemented for competitive systems described by the LV equations. This formalism allowed for introducing non-commutative features into prey-predator systems and studying its effects in the phase-space.

Departing from a classical LV Hamiltonian pattern for the number of species, $y = e^{-x}$ and $z = e^{-k}$, with $[x,\,k] = 0$, and subsequently considering the non-commutative algebra, $[x,\,k] = i$, the quantum features of the LV system have been studied.
The quantum analogue hypothesis, $[x,\,k] = i$, means that simultaneous measurements of $x$ and $k$ have no additional constraint(s) besides the Hamiltonian one. In this context, the fluctuations of the number of individuals, implicitly described by $\delta x$ and $\delta k$, follow a non-deterministic evolution parameterized as $\delta x\,\delta k \gtrsim 0$, expressed by Wigner currents, which drive the statistical and probability effects over the quantum ensembles. 
In this quantum-like scenario, $\delta x\,\delta k \gtrsim 0$ suggests a minimal phase-space volume unity (estimated by $ \hbar$ or by another effective parameter, in case of an effective quantum-like scenario), which suppresses definitive extinction scenarios.

Critical transitions related to the shift between classical and quantum regimes were identified by the presence of topological effects over the phase-space and quantified in terms of stationarity and Liouvillianity measures.
Macroscopically interpreted as sudden and unexpected changes in the state of a competitive LV system, such transitions are usually associated with environmental changes which, in case of our analysis, could be driven by TD and gaussian ensembles.
In fact, for parametrical proposals involving fitness functions, creation of hierarchies and quantum mixing \cite{Anna}, gaussian quantum ensembles were shown to be the more suitable Hilbert space state configuration for comparing quantum to classical regimes.

In the framework here considered, the emergence of quantum patterns and classical to quantum transitions do not involve any obvious loss of stability. However, once detected, this was stratified into subclassifications provided by the hyperbolic stability criteria.
For the phase-space oscillating population portraits interpreted as the {\it quantum analog} of a prey-predator-like dynamics (cf. Hamiltonian Eq.~\eqref{Ham}), one notices that for anisotropic population oscillation patterns ($a\neq 1$), quantum corrections affect the stability pattern.
In particular, for $a < 1$ and $\alpha \neq 0$, the quantum effects asymptotically drive the system to periodic population extinction and revivals. Such a result approaches the one from Ref.~\cite{PRE-LV} where discreteness of the populations destroy the mean-field stability and eventually drive the system toward extinction of one or both species.
Conversely, for $a > 1$ and $\alpha \neq 0$, the population oscillations are asymptotically suppressed, i.e. they approximately stabilize around equilibrium points. Interestingly, such a result approaches the so-called {\it extinction of oscillating populations} from Ref.~\cite{PRE-LV2}, as well as exhibits similar patters to those depicted by Ref.~\cite{PRE-LV4} where a stochastic version of the Rosenzweig-MacArthur predator-prey model \cite{Mac63} is considered.

Finally, also for gaussian ensembles, a subliminal effect was investigated. Even for a non-Liouvillian pattern, with $\mbox{\boldmath $\nabla$}_{\xi} \cdot \mathbf{w} \neq 0$, quantum ensembles may also exhibit stationary evolution patterns, i.e. with $\mbox{\boldmath $\nabla$}_{\xi} \cdot \mbox{\boldmath $\mathcal{J}$} = 0$, exhibiting a kind of {\it camouflage} effect of quantum distortions coupled to non-linearity.

To summarize, our analysis is an additional step towards a description of quantum effects and their associated stability issues in competitive LV systems. For all these reasons, one can interpret the framework considered here as a novel tool for unveiling the origin of critical transitions and collective effects in the context of non-linear Hamiltonian systems which admit some quantization hypothesis.

\vspace{.5 cm}
{\it Acknowledgments -- The work of AEB is supported by the Brazilian Agency FAPESP (Grant No. 2020/01976-5).}

\end{document}